\documentclass[12pt,letterpaper,a4paper]{article} 

\usepackage[includeheadfoot,
            marginratio={1:1,1:1}, 
            width=440pt, 
            height=688pt,]{geometry}

\usepackage{amsmath}
\usepackage{amsfonts}
\usepackage{amssymb}
\usepackage{stmaryrd}
\usepackage{dsfont}
\usepackage{graphicx}
\usepackage{amsthm,amscd}
\usepackage{extarrows}
\usepackage{hyperref}
\usepackage[all]{hypcap}

\newtheoremstyle{note} 
  {5pt}                
  {5pt}                
  {}                   
  {\parindent}         
  {\bfseries}          
  {.}                  
  {.5em}               
  {}                   

\newtheorem*{FR}{Field redefinition}
\newtheorem*{dual}{$O(d,d)$-duality}
\theoremstyle{definition}

\newcommand{\nc}{\newcommand}
\nc{\lb}{\llbracket}
\nc{\rb}{\rrbracket}
\nc{\gl}{\llbracket}
\nc{\gr}{\rrbracket}

\newcommand{\eq}[1]{\begin{equation}
                     \begin{split} #1 \end{split}
                     \end{equation}}

\allowdisplaybreaks[2]
\numberwithin{equation}{section}

\begin{document}

\vspace*{-1.5cm}

\begin{flushright}
  {\small
  MPP-2014-68
  }
\end{flushright}

\vspace{1.5cm}


\begin{center}
{\LARGE
O(d,d)-Duality in String Theory
}
\end{center}


\vspace{0.4cm}

\begin{center}
Felix Rennecke$^{1}$
\end{center}


\vspace{0.4cm}

\begin{center} 
\emph{$^{1}$ Max-Planck-Institut f\"ur Physik (Werner-Heisenberg-Institut), \\ 
   F\"ohringer Ring 6,  80805 M\"unchen, Germany }
\end{center} 

\vspace{0.4cm}


\begin{abstract}
A new method for obtaining dual string theory backgrounds is presented. Preservation of the Hamiltonian density and the energy momentum tensor induced by $O(d,d)$-transformations leads to a relation between dual sets of coordinate one-forms accompanied by a redefinition of the background fields and a shift of the dilaton. The necessity of isometric directions arises as integrability condition for this map. The isometry algebra is studied in detail using generalised geometry. In particular, non-abelian dualities and $\beta$-transformations are contained in this approach. The latter are exemplified by the construction of a new approximate non-geometric background.
\end{abstract}


{\small
\tableofcontents
}

\section{Introduction}
\label{sec:intro}

Dualities are the manifestation of the rich symmetry structure distinctive of string theory. By virtue of relating seemingly different string theory solutions they proved to be a valuable guide for finding new phenomena such as D-branes, mirror symmetry or exotic solutions. In particular, string theory compactified on a $d$-dimensional torus admits the T-duality group $O(d,d;\mathbb{Z})$ (a review is found in \cite{Giveon:1994fu}). In this vein, double field theory \cite{Siegel:1993xq,Siegel:1993th,Hull:2009mi,Hull:2009zb,Hohm:2010jy,Hohm:2010pp} (recent reviews are found in \cite{Aldazabal:2013sca,Berman:2013eva,Hohm:2013bwa}) aims for a manifest $O(d,d)$-invariant formulation of string theory, although evidence for duality beyond abelian duality for toroidal backgrounds is scarce.

The conventional approach to (non-abelian) T-duality relies on the existence of a consistent gauged sigma model associated with isometries \cite{Rocek:1991ps,delaOssa:1992vc} (a Hamiltonian approach is found in \cite{Alvarez:1994wj}). In the case of abelian isometries, the gauged sigma model identifies seemingly different theories as being equivalent in accordance with the Buscher rules \cite{Buscher:1987sk,Buscher:1987qj}. However, for non-abelian isometries, the theories connected by the gauged sigma model are  in general not equivalent \cite{Alvarez:1993qi,Giveon:1993ai}. This problem arises with the introduction of a gauge field which possibly admits non-trivial holonomies as well as the appearance of anomalies \cite{Hull:1989jk,Alvarez:1993qi}.

Circumventing some of these difficulties and extending duality to more general, non-constant $O(d,d)$-transformations allows for a deeper understanding of the symmetry structure of string theory.\footnote{See for example \cite{Andriot:2009fp,Andriot:2010ju} for applications.}
To pursue in this direction, the following observation is made.
The background dependent description of string theory by a two-dimensional sigma model naturally gives rise to the indefinite orthogonal group $O(d,d)$ as it preserves one component of the world-sheet energy momentum tensor.
Induced mappings of the (pulled-back) coordinate one-forms $dX^a$ and redefinitions of the background fields leave the Hamiltonian density and the remaining component of the energy momentum tensor invariant as well. Therefore, the mapping of coordinate one-forms together with the background redefinition represents a classical duality. The basic properties of this duality, which includes (non-abelian) T-duality and $\beta$-transformations, are explored in this paper.

In particular, the initial coordinate one-form $dX^a$ is mapped to a dual form $d\widetilde{X}^a$ whose integrability implies the existence of certain isometries of the background. The isometry algebra is formulated in terms of Lie algebroids (see \cite{Blumenhagen:2012pc,Blumenhagen:2012nt,Blumenhagen:2013aia} for applications to string theory) and a twisted Courant bracket \cite{Severa:2001qm}.\footnote{See \cite{Alekseev:2004np,Plauschinn:2013wta} for applications of the bracket to isometries.} Closure of the algebra gives rise to the conditions required for a consistent gauging of the isometries \cite{Hull:1989jk} and therefore connects the present approach to the conventional method. The problem of anomalies is absent in the suggested procedure, but the conditions for anomaly freedom can be retrieved from a Dirac structure for the isometry algebra \cite{Alekseev:2004np}.

Having received little attention in the literature, $\beta$-transformations are of particular interest here.\footnote{In the context of non-geometric backgrounds they are discussed in \cite{Andriot:2013xca,Andriot:2014uda}, in the context of AdS/CFT they appeared for example in \cite{Minasian:2006hv} and a relation to Ehlers transformations in heterotic string theory can be found in \cite{Kumar:1995np,Biswas:1995cp}.} They induce classical duality, if $\beta$ is a Poisson structure. This \emph{Poisson-duality} is applied to the rectangular three-torus with constant $H$-flux. Moreover, for backgrounds related by $\beta$-transformations to be conformal, an appropriate shift of the dilaton is deduced.

The paper is organised as follows. In section~2 classical features of the string sigma model and T-duality are recapitulated. In particular, the appearance of $O(d,d)$ is extracted from the constraints in a Hamiltonian formulation. Section~3 is devoted to the detailed discussion of $O(d,d)$-duality. It includes the study of the integrability conditions for the mapping of coordinate one-forms manifest in the isometry algebra, the main elements of $O(d,d)$ and the special role of the dilaton for duality on the quantum level. The section closes with an example providing a new approximate non-geometric background.


\section{The bosonic string sigma model}
String theory is described in a background dependent fashion by a two-dimensional non-linear sigma model. For discussing closed bosonic strings, $\Sigma$ is a two-dimensional manifold with metric $h=\mathrm{diag}(-1,1)$ and $\partial\Sigma=\emptyset$. The world-sheet $\Sigma$ is embedded into a $d$-dimensional Riemannian manifold $M$ via $X:\Sigma\hookrightarrow M$. Having coordinates $\{x^a\}_{a=1}^d$ for $M$, their pull-back to $\Sigma$ is denoted $X^a=X^*x^a$. With $\star$ the Hodge operator with respect to $h$, the action can be written as\footnote{The conventions are as follows: The coordinates on $\Sigma$ are $\{\tau,\sigma\}$ and the orientation is given by the volume element $d\tau\wedge d\sigma$. Then the Hodge operator is given by $\alpha\wedge\star\beta = h(\alpha,\beta) d\tau\wedge d\sigma$ for arbitrary $\alpha,\beta\in\Gamma(\Lambda^nT^*\Sigma)$. For a decomposition $\alpha=\alpha_1\wedge\dots\wedge\alpha_n$ and similarly for $\beta$ the insertion into the metric is defined as $h(\alpha,\beta)=\det h(\alpha_i,\beta_j)$.}
\eq{
\label{sigma}
	S(X;G,B)=\frac{1}{4\pi\alpha'}\int_\Sigma\left[G(X)_{ab}\,dX^a\wedge\star dX^b 
		+ B(X)_{ab}\,dX^a\wedge dX^b\right]\,.
}
$G$ is a Riemannian metric on the target-space $M$ and $B$ a two-form; the pair $(G,B)$ will be called the \emph{background}. Moreover, $d$ denotes the exterior derivative on $T\Sigma$ while $\mathbf{d}$ denotes the exterior derivative on $TM$. The dilaton will be discussed separately in section~\ref{sec:conform}.
The immediate classical features of \eqref{sigma} are the following.
\begin{itemize}
\item Varying the action with respect to $X^a$ yields the equation of motion
\eq{\label{eom:X}
	d\star dX^a + \Gamma^a{}_{bc}\,dX^b\wedge\star dX^c = \tfrac{1}{2}\,G^{am}\,H_{mbc}\,dX^b\wedge dX^c
}
with $H_{abc}$ the components of $H=\mathbf{d}B$ and $\Gamma^a{}_{bc}=\frac{1}{2}G^{am}(\partial_bG_{mc}+\partial_cG_{mb}-\partial_mG_{bc})$ the coefficients of the Levi-Civita connection on $TM$. Possible boundary terms are neglected. For $H=0$ \eqref{eom:X} is the generalization of the geodesic equation for a world-sheet. In the presence of the $H$-term, \eqref{eom:X} can be interpreted as geodesic motion of a membrane in Einstein-Cartan theory with Bismut connection $\Gamma^{a}{}_{bc}-\frac{1}{2}G^{am}H_{mbc}$.
\item The equation of motion for a general world-sheet metric $h$ is vanishing of the energy-momentum tensor, $T_{\alpha\beta}=0$. In the conformal gauge chosen here, this has to be considered as constraints which read
\eq{\label{Tcons}
	G_{ab}(\partial_\tau X^a\partial_\tau X^b+\partial_\sigma X^a\partial_\sigma X^b) =0
	\quad\quad\&\quad\quad
	G_{ab}\,\partial_\tau X^a\partial_\sigma X^b=0\,.
}
Hence the dynamics of the theory is determined by the equation of motion \eqref{eom:X} accompanied with the constraints \eqref{Tcons}.
\end{itemize}
In the following the Hamiltonian description will be discussed briefly.

\subsection{Hamiltonian description}
The Hamiltonian density can be determined from the Lagrangian density in \eqref{sigma} by performing a Legendre transformation with respect to the canonical momentum and $\tau$-derivative of the coordinate fields $X^a$. In principle there are two possibilities for canonically conjugate variables to the coordinate field $X^a$, which will become important for the discussion of duality:
\begin{itemize}
\item the \emph{canonical momentum} $P_a=\frac{\partial L}{\partial\partial_\tau X^a}=\frac{1}{2\pi\alpha'}(-G_{ab}\partial_\tau X^b +B_{ab}\partial_\sigma X^b)$,
\item the \emph{canonical winding} $W_a=\frac{\partial L}{\partial\partial_\sigma X^a}=\frac{1}{2\pi\alpha'}(G_{ab}\partial_\sigma X^b -B_{ab}\partial_\tau X^b)$.
\end{itemize}
However, by virtue of the first constraint in \eqref{Tcons}, the Hamiltonian density arising from a Legrendre transformation with respect to $P$ and $\partial_\tau X$ coincides with the one resulting from a transformation with respect to $W$ and $\partial_\sigma X$ since
\eq{\label{xpxw}
	\partial_\tau X^a\,P_a = \partial_\sigma X^a\,W_a.
}
Performing the transformation, the Hamiltonian density can be written as
\eq{\label{ham}
	\mathrm{Ham}(X;G,B)&=-\frac{1}{4\pi\alpha'}\begin{pmatrix}\partial_\sigma X\\2\pi\alpha'P\end{pmatrix}^t
	\mathcal{H}(G,B)
	\begin{pmatrix}\partial_\sigma X\\2\pi\alpha'P\end{pmatrix} \\
	&=\frac{1}{4\pi\alpha'}\begin{pmatrix}\partial_\tau X\\-2\pi\alpha'W\end{pmatrix}^t
	\mathcal{H}(G,B)
	\begin{pmatrix}\partial_\tau X\\-2\pi\alpha'W\end{pmatrix}\,,
}
where the \emph{generalised metric}
\eq{\label{H}
	\mathcal{H}(G,B)=\begin{pmatrix} G-BG^{-1}B & BG^{-1}\\ -G^{-1}B & G^{-1}\end{pmatrix}
}
is introduced.
Defining the generalised vectors
\eq{\label{APW}
	A_P(X)&=\partial_\sigma X^a\,\frac{\partial}{\partial x^a} + 2\pi\alpha'\, P_a\, \mathbf{d}x^a \\
	A_W(X)&=\partial_\tau X^a\,\frac{\partial}{\partial x^a}  -2\pi\alpha'\, W_a\, \mathbf{d}x^a
}
in $TM\oplus T^*M$, the Hamiltonian density \eqref{ham} is proportional to the squared length of $A_P$ and $A_W$ as measured by the generalised metric \eqref{H}: $\mathrm{Ham}=-\frac{1}{4\pi\alpha'}||A_P||_{\mathcal{H}}^2$.

\subsection{Appearance of $O(d,d)$}\label{subsec:odd}
Already on the classical level the indefinite orthogonal group $O(d,d)$ appears naturally. In terms of the generalised vector $A_P$ \eqref{APW}, the constraints \eqref{Tcons}, i.e. the components of the energy momentum tensor can be rewritten as
\eq{\label{classical_cons}
	A_P^t\,\mathcal{H}(G,B)\,A_P=0 \quad\quad\&\quad\quad
	A_P^t\,\eta\, A_P=0\,.
}
As the first constraint sets the Hamiltonian density to zero, the constrained dynamics is completely governed by \eqref{xpxw}. For the second constraint we have introduced the matrix
\eq{\label{eta}
	\eta = \begin{pmatrix} 0 & \mathds{1}\\ \mathds{1}&0\end{pmatrix}\,,
}
which defines the group $O(d,d)$: A $d\times d$-matrix $\mathcal{T}$ is an element of $O(d,d)$ if and only if
\eq{
	\mathcal{T}^t\,\eta\,\mathcal{T}=\eta\,,
}
i.e. if it leaves the matrix $\eta$ invariant. In particular, the generalised metric \eqref{H} is an element of $O(d,d)$ and the inverse is given generally by
\eq{\label{oddinv}
	\mathcal{T}^{-1} = \eta\,\mathcal{T}^t\,\eta \quad\quad\forall\mathcal{T}\in O(d,d).
}
Therefore all admissible generalised vectors solving the second constraint in \eqref{classical_cons} are related by an $O(d,d)$-transformation via $A_P' =\mathcal{T}A_P$. For $A_P'$ to solve the first constraint as well, a compensating $O(d,d)$-conjugation with $\mathcal{T}^{-1}$ has to be applied to the generalised metric \eqref{H}. This transformation will be the subject of the duality discussed in the next section.

\subsection{Review of T-duality}\label{sec:Trev}
The conventional procedure for obtaining T-dual sigma models by gauging isometries will be reviewed briefly \cite{Buscher:1987sk,Rocek:1991ps}. For simplicity, a single isometry of \eqref{sigma} generated by a vector field $k$ is considered. In the case of multiple non-abelian isometries the gauging procedure can be found in \cite{Hull:1989jk}. With respect to the infinitesimal coordinate transformation
\eq{\label{tdiff}
	X^a\to X^a+\epsilon\,k^a
}
the sigma model \eqref{sigma} transforms as $S\to S+\delta S$ with
\eq{
	\delta S(X;G,B)= \frac{\epsilon}{4\pi\alpha'}\int_\Sigma
		\left[(L_kG)_{ab}\,dX^a\wedge\star dX^b + (L_kB)_{ab}\,dX^a\wedge dX^b\right] \,.
}
$L_k$ denotes the Lie derivative along the vector field $k$.
Thus $k$ generates an isometry of the sigma-model if it satisfies\footnote{Note that $(L_kB)_{ab}\,dX^a\wedge dX^b=X^*(L_kB)$ and $X^*(\mathbf{d}\nu)=d(X^*\nu)$ for $\nu\in\Gamma(T^*M)$.}
\eq{\label{iso_cond}
	L_kG=0 \quad\quad\&\quad\quad L_kB = \mathbf{d}\nu \quad\mathrm{for}\,\nu\in\Gamma(T^*M)\,.
}
By using that a gauge transformation $B\to B+\mathbf{d}\omega$ induces the transformation $\nu\to\nu+L_k\omega$, a gauge in which $\nu=0$ can be found. Assuming this gauge to be chosen in adapted coordinates $k=\frac{\partial}{\partial X^1}$ allows to gauge the isometry generated by $k$ via minimal coupling: Introducing the gauge field $A\in\Gamma(T^*\Sigma)$ which transforms under the local version of \eqref{tdiff} as $\delta A = -d\epsilon$
, minimal coupling amounts to the substitution $dX^1\to DX^1=dX^1+A$. 
Choosing the gauge $A\to A-dX^1$, the gauged sigma model takes the form $S_\mathrm{gauged}=S(X^m;G,B)+S_g$ with
\eq{
	S_g= \frac{1}{4\pi\alpha'}\int_\Sigma
	\bigl(G_{11}\,A\wedge\star A +2\,G_{1m}\,A\wedge\star dX^m + 2\,B_{1m}\,A\wedge dX^m
	-2\,A\wedge d\lambda\bigr)
}
for $m\neq 1$. Integrating out the Lagrange multiplier $\lambda$ yields $A=dX^1$ locally and gives back the initial sigma model \eqref{sigma}. Integrating out the gauge field gives
\eq{\label{eom:A}
	\star A = -\frac{1}{G_{11}}\bigl(G_{1m}\,\star\!dX^m +B_{1m}\,dX^m-d\lambda\bigr)\,.
}
Plugging this back into the gauged action and considering $d\lambda=d\widetilde{X}^1$ as a new coordinate, the resulting action can be written as \eqref{sigma} with the new background $(g,b)$ given by the \emph{Buscher rules} \cite{Buscher:1987sk}
\eq{\label{buscher}
	g_{11}=\frac{1}{G_{11}}\,, \quad\quad g_{1m}=-\frac{B_{1m}}{G_{11}}\,,
	\quad&\quad g_{mn}=G_{mn}-\frac{G_{m1}G_{1n}+B_{m1}B_{1n}}{G_{11}}\,, \\
	b_{1m}=-\frac{G_{1m}}{G_{11}}\,,
	\quad&\quad b_{mn}=B_{mn}-\frac{G_{m1}B_{1n}+B_{m1}G_{1n}}{G_{11}}\,.
}
Hence, T-duality can be performed along the direction of an isometry and the dual backgrounds are related by \eqref{buscher}. T-duality also introduces a new coordinate one-form $d\widetilde{X}$ which can be related to $dX^1$ on-shell by \eqref{eom:A}: Identifying $A=dX^1$ and $d\lambda=d\widetilde{X}^1$, \eqref{eom:A} can be written as
\eq{\label{Tdual_coord}
	d\widetilde{X}^1 = G_{1a}\star\!dX^a +B_{1a}dX^a\,.
}
This is the conserved current associated to the isometry \eqref{tdiff} generated by $k=\frac{\partial}{\partial X^1}$. For the gauging of \eqref{sigma} to be consistent, the global structure of the world-sheet $\Sigma$ has to be taken into account \cite{Rocek:1991ps,Alvarez:1993qi}. In particular, the gauge field $A$ can have non-trivial holonomies. For gauging multiple isometries $\{k_i\}$, further conditions apart from \eqref{iso_cond} arise \cite{Hull:1989jk}: With $\kappa_i=\nu_i-\iota_{k_i}B$ such that $\iota_{k_i}H=\mathbf{d}\kappa_i$ and $[k_i,k_j]=F^m{}_{ij}k_m$, also
\eq{\label{cons_gauge}
	L_{k_i}\kappa_j=F^m{}_{ij}\,\kappa_m \quad\quad\mathrm{and}\quad\quad
	\iota_{k_i}\kappa_j + \iota_{k_j}\kappa_i=0
}
have to be satisfied. The second condition ensures the gauged sigma model to be free of anomalies.
In the next chapter a different approach to duality is developed and the Buscher rules \eqref{buscher} with \eqref{Tdual_coord} as well as the conditions \eqref{iso_cond}, \eqref{cons_gauge} are encountered as special cases.

\section{$O(d,d)$-duality}
In this section a new way of performing duality is proposed by redefining the background and identifying dual coordinates directly. This avoids the procedure of gauging and accordingly circumvents the problem of anomalies. 
As observed in section~\ref{subsec:odd} the admissible generalized vectors \eqref{APW} satisfying the constraints \eqref{classical_cons} are related by $O(d,d)$-transformations, which implies a simultaneous inverse transformation of the generalized metric \eqref{H}. This, in turn, leaves the Hamiltonian density \eqref{ham} and the energy momentum tensor \eqref{classical_cons} invariant. This duality will be described in detail in the following.

\subsection{Field redefinitions and duality}
\label{sec:odd_details}
The admissible generalized vector $A_P$ will be transformed by\footnote{The bar over the index indicates the one associated to the domain. As to operations with linear maps, inversion swaps indices (e.g. $t_{11}^{-1}\equiv (t_{11})^{\bar a}{}_a:TM\to TM$) and transposition commutes them (e.g. $t_{11}^t\equiv (t_{11})_{\bar a}{}^a:T^*M\to T^*M$). The combination $f^{-t}=(f^{-1})^t$ is used as well.}
\eq{\label{odd_comp}
	\mathcal{T}=\begin{pmatrix}t_{11}&t_{12}\\t_{21}&t_{22}\end{pmatrix}\,\in O(d,d)
	\quad\quad\mathrm{with}\quad
	\arraycolsep2pt
	\begin{array}{ccccccc}
		t_{11}&\equiv& (t_{11})^a{}_{\bar a} &:&TM&\to& TM \\
		t_{12}&\equiv&(t_{12})^{a\bar a}&:&T^*M&\to& TM \\
		t_{21}&\equiv&(t_{21})_{a\bar a}&:&TM&\to& T^*M \\
		t_{22}&\equiv&(t_{22})_a{}^{\bar a}&:&T^*M&\to& T^*M
	\end{array}
}
as $A_P\to\mathcal{T}^{-1}A_P$. In order for the first constraint in \eqref{classical_cons} to remain satisfied the generalized metric has to be conjugated with $\mathcal{T}$ simultaneously:
\eq{\label{Htrafo}
	\mathcal{H}(G,B)\to\mathcal{T}^t\,\mathcal{H}(G,B)\,\mathcal{T}\equiv\mathcal{H}(g,b)\,.
}
By \eqref{ham}, this simultaneous transformation leaves the Hamiltonian density invariant, which may give an equivalent theory. This specific equivalence will be called \emph{$\mathcal{T}$-duality} and will be explored in the rest of the paper. In \eqref{Htrafo}, $\mathcal{H}(g,b)$ refers to a redefinition of the background in order for the generalized metric to have the standard form \eqref{H} as follows.
\begin{FR}[\cite{Blumenhagen:2013aia}]
An $O(d,d)$-rotated generalized metric $\mathcal{T}^t\mathcal{H}(G,B)\mathcal{T}$ takes the standard form \eqref{H} with respect to the new background $(g,b)$. In terms of the automorphism\footnote{Invertibility of $\gamma$ was shown in the appendix of \cite{Blumenhagen:2013aia}. In particular, the target-space metric $G$ being positive definite is a sufficient condition.}
\eq{\label{gamma}
	\gamma = t_{22} +(G-B)t_{12}: T^*M\to T^*M
}
the new background $(g,b)$ is given by
\eq{\label{redef}
	g = \gamma^{-1}\,G\,\gamma^{-t} \quad\quad\&\quad\quad
	b  = \gamma^{-1}(\gamma\,\delta^t-G)\gamma^{-t}
}
with $\delta = t_{21}+(G-B)t_{11}$.
\end{FR}\noindent
In \cite{Blumenhagen:2013aia} the field redefinition \eqref{redef} was used to study the geometric structure of the target space low energy effective theory for \eqref{sigma}. In particular, \eqref{gamma} induces an anchor for a Lie algebroid describing the associated geometry and gauge theory.

The simultaneous rotation of the generalized vectors \eqref{APW} gives rise to redefined phase space coordinates.
They can be read-off from the transformation
\eq{
	A_P\to\mathcal{T}^{-1}\,A_{P}(X)\equiv A_{\widetilde{P}}(\widetilde{X})
	=\begin{pmatrix} \partial_\sigma\widetilde{X} \\ 2\pi\alpha'\widetilde{P}\end{pmatrix}
}
and analogously for the winding vector. Using \eqref{oddinv} the dual pair becomes
\eq{\label{dual_pair}
	\partial_\sigma\widetilde{X}^{\bar a}&= -(t_{12})^{\bar am}G_{ma}\,\partial_\tau X^a
		+\bigl[(t_{22})^{\bar a}{}_a
		+(t_{12})^{\bar am}B_{ma}\bigr]\partial_\sigma X^a\,, \\
	\widetilde{P}_{\bar a} &= \tfrac{1}{2\pi\alpha'}\left\{-(t_{11})_{\bar a}{}^mG_{ma}\,\partial_\tau X^a
		+\bigl[(t_{21})_{\bar aa}
		+(t_{11})_{\bar a}{}^mB_{ma}\bigr]\partial_\sigma X^a\right\}\,.
}
For determining the dual coordinates the $\tau$-derivative of $X^a$ is required as well.
In principle, $\partial_\tau\widetilde{X}^{\bar a}$ can be computed from the general equation for the dual momentum $\widetilde{P}$. Since the Hamiltonian densities with respect to momentum and winding coincide by \eqref{xpxw}, it is easier to deduce it directly from the winding vector $A_{\widetilde{W}}(\widetilde{X})$ as above:
\eq{\label{dual_vel}
	\partial_\tau\widetilde{X}^{\bar a} 
	& = -(t_{12})^{\bar am}G_{ma}\,\partial_\sigma X^a
		+\bigl[(t_{22})^{\bar a}{}_a
		+(t_{12})^{\bar am}B_{ma}\bigr]\partial_\tau X^a\,,\\
	\widetilde{W}_{\bar a}&=\tfrac{-1}{2\pi\alpha'}\left\{-(t_{11})_{\bar a}{}^mG_{ma}\,\partial_\sigma X^a
		+\bigl[(t_{21})_{\bar aa}
		+(t_{11})_{\bar a}{}^mB_{ma}\bigr]\partial_\tau X^a\right\}\,.
}
Having determined both world-sheet derivatives of the dual coordinates\footnote{Using the relations between the elements of the $O(d,d)$-matrix $\mathcal{T}^{-1}$, \eqref{dual_pair} and \eqref{dual_vel} satisfy the constraint $\partial_\tau\widetilde{X}^{\bar a}\widetilde{P}_{\bar a} = \partial_\sigma\widetilde{X}^{\bar a}\widetilde{W}_{\bar a}$ \eqref{xpxw} as well.} $\widetilde{X}^{\bar a}$, the main result of this paper can be formulated.
\begin{dual}
Let $\{e_a\}_{a=1}^d$ be a frame for $TM$ and $\{e^a\}_{a=1}^d$ its dual.
For $\mathcal{T}\in O(d,d;\mathcal{C}^\infty(M))$, the sigma model $S(X;G,B)$ \eqref{sigma} is $\mathcal{T}$-dual to $S(\widetilde{X};g,b)$ on-shell with the coordinates related via
\eq{\label{dual_coord}
	d\widetilde{X}^{\bar a} = \bigl[(t_{12})^{\bar am}G_{ma}\bigr]\star\! dX^a
		+\bigl[(t_{22})^{\bar a}{}_a+(t_{12})^{\bar am}B_{ma}\bigr] dX^a
}
and the backgrounds related by the field redefinition \eqref{redef}, provided
\eq{\label{integ}
	L_{t_{12}^\sharp e^{\bar a}}G =0 \quad\quad\mathrm{and}\quad\quad
	L_{t_{12}^\sharp e^{\bar a}}B = -\mathbf{d}\bigl(t_{22}^\sharp e^{\bar a}\bigr) \,.
}
Here $t_{12}^\sharp e^{\bar a} = (t_{12})^{\bar am}e_m$ and $t_{22}^\sharp e^{\bar a} = (t_{22})^{\bar a}{}_me^m$. The requirement \eqref{integ} is the integrability condition for \eqref{dual_coord}.
\end{dual}\noindent
Equation \eqref{dual_coord} is the combination of \eqref{dual_pair} and \eqref{dual_vel}. The integrability condition \eqref{integ} can be deduced by differentiating \eqref{dual_coord} and using the equations of motion \eqref{eom:X} as well as $\iota_vH = L_v B -d\iota_v B$ for any vector field $v$. Thus, in particular, the duality is only valid on-shell. Further restriction arise from the algebra spanned by the vectors $t_{12}^\sharp e^{\bar a}$, which will be discussed  in section~\ref{sec:iso_alg}.

$O(d,d)$-duality can be described in terms of the \emph{duality map} as follows.
By defining $dX=dX^ae_a\in\Gamma(TM\otimes T^*\Sigma)$, the duality automorphism
\eq{
	\mathfrak{D}:\Gamma(TM\otimes T^*\Sigma)\to\Gamma(TM\otimes T^*\Sigma)\,;\,\,
	dX\mapsto d\widetilde{X}=\mathfrak{D}(dX)
}
follows from \eqref{dual_coord}. In matrix notation it can be written globally as
\eq{\label{dualitymap}
	\mathfrak{D}=t_{12}^t\,G\otimes\star+\bigl(t_{22}^t+t_{12}^t\,B\bigr)\otimes\mathrm{id}_{T^*\Sigma}\,.
}
Indeed, the inverse of the duality map \eqref{dualitymap} can be easily determined by the inverse procedure and reads
\eq{\label{dualitymap_inv}
	\mathfrak{D}^{-1}=t_{12}\,g\otimes\star+\bigl(t_{11}+t_{12}\,b\bigr)\otimes\mathrm{id}_{T^*\Sigma}
}
in terms of the dual background \eqref{redef}. Hence $O(d,d)$-duality is invertible. The subsection is closed with the following observations and remarks.

\subsubsection*{Duality and isometries}
The dual coordinates \eqref{dual_coord} and the integrability conditions \eqref{integ} can be interpreted as follows. As can be seen by comparing \eqref{integ} with \eqref{iso_cond}, the integrability condition ensures the infinitesimal target space diffeomorphism generated by the vector field $t_{12}^\sharp e^{\bar a}$ to be an isometry of \eqref{sigma}. The one-form $\nu$ in \eqref{iso_cond} is explicitly determined to read $\nu = -t_{22}^\sharp e^{\bar a}$ up to exact terms. These special isometries will be called \emph{duality isometries} in the following. It can be checked that the dual coordinates $d\widetilde{X}^{\bar a}$ coincide with the conserved current $J^a$ associated to the isometry $X^a\to X^a+\epsilon\, t_{12}^\sharp e^a$. In particular, the duality map \eqref{dualitymap} interchanges the $TM$-valued coordinate one-forms $dX$ with the $TM$-valued conserved currents $J=d\widetilde{X}$.

\subsubsection*{Is \eqref{dual_coord} a coordinate transformation?}
By using the Poincar\'e lemma and the integrability conditions \eqref{integ}, \eqref{dual_coord} is locally exact. Then the local primitive for $d\widetilde{X}^{\bar a}$ might be interpreted as dual pulled-back coordinate $\widetilde{X}^{\bar a}$. First, this raises the question whether the coordinates on the target-space are changed, i.e. $\widetilde{X}^{\bar a}=X^*(\tilde{x}^{\bar a})$, or the embedding is changed, i.e. $\widetilde{X}^{\bar a}=\widetilde{X}^*(x^{\bar a})$. Second, it is not clear if the resulting relation $\widetilde{X}^{\bar a}(X)$ is invertible, i.e. if $X^a(\widetilde{X})$ can be found. In particular, both questions are important for the interpretation of the field redefinition \eqref{redef} due to \eqref{Htrafo}, since the new background still depends on the initial coordinates.\footnote{I thank the referee for pointing out this problem of interpretation.} This also effects the interpretation of \eqref{dualitymap_inv}.

In the case of constant $O(d,d)$-transformations and constant backgrounds, \eqref{dual_coord} can be integrated directly and the relation between the dual coordinates is invertible: The equations of motion \eqref{eom:X} reduces to the wave equation and is solved by $X^a(\tau,\sigma)=X^a_+(\sigma^+)+X^a_-(\sigma^-)$ with the light-cone coordinates $\sigma^\pm=\tau\pm\sigma$. Using that $O(d,d)$-duality with respect to the unit matrix leaves everything invariant, \eqref{dual_coord} can be integrated to give
\eq{\label{const_dual}
	\widetilde{X}^{\bar a}_+ &= \bigl[(t_{22})^{\bar a}{}_a +(t_{12})^{\bar am}(B_{ma}-G_{ma})\bigr]X^a_+ \\
	\widetilde{X}^{\bar a}_- &= \bigl[(t_{22})^{\bar a}{}_a +(t_{12})^{\bar am}(B_{ma}+G_{ma})\bigr]X^a_-\,.
}
Invertibility of $t_{22}+t_{12}(B\pm G)$ is equivalent to the invertibility of \eqref{gamma}. Thus in this case \eqref{dual_coord} gives rise to a proper change of coordinates. Keeping the necessity of a positive definite metric for invertibility of \eqref{gamma} in mind, this shows that $O(d,d)$-duality includes the well-known case of the T-duality group $O(d,d;\mathbb{Z})$ for toroidal target-spaces; the transformations have to be integer in order for the periodicities to be preserved (see e.g. \cite{Giveon:1994fu}).\footnote{See sections~\ref{sec:iso_alg} and \ref{sec:proto} for further relations to the known cases.} The novelty is that $O(d,d)$-duality gives the precise relation between the dual coordinates.

For the more general case of non-constant backgrounds or non-constant $O(d,d)$-transformation the question about invertibility remains open. 

\subsubsection*{Comment on global issues}
In the conventional approach to duality by gauging the isometries, global issues might prevent the "dual" theories from being dual. They are related to the possibility of having non-trivial holonomies for the newly introduced gauge fields \cite{Rocek:1991ps,Alvarez:1993qi,Giveon:1993ai,Giveon:1994mw}. This discussion takes place at the level of the gauged sigma model and cannot be repeated here. In particular, $O(d,d)$-duality gives equivalent classical theories by construction.\footnote{Upon taking all the consistency requirements into account.} However, the global structure of the dual space is determined by the winding number of $d\widetilde{X}^{\bar a}$
\eq{
	c_\mathrm{wind}\bigl(\widetilde{X}^{\bar a}\bigr)=\oint_\gamma d\widetilde{X}^{\bar a}\,,
}
with $\gamma$ a closed curve in $\Sigma$. This is related to the winding number of the initial coordinate one-forms by \eqref{dual_coord}. 

\subsection{A Lie algebroid for duality isometries and consistency}\label{sec:iso_alg}
$O(d,d)$-duality is feasible if $t_{12}^\sharp e^{\bar a}$ generates the isometries. Moreover, the isometry algebra has to close and has to satisfy the Jacobi identity. For their part, Killing vector fields are closed: $[t_{12}^\sharp e^{\bar a},t_{12}^\sharp e^{\bar b}]$ is a Killing vector field as well. However, closed duality isometries require $[t_{12}^\sharp e^{\bar a},t_{12}^\sharp e^{\bar b}]$ to be a linear combination of the generators $t_{12}^\sharp e^{\bar a}$. In this section the consistency of the isometry algebra is investigated, which can be described in terms of Lie algebroids \cite{Mackenzie05}.

The vector fields $t_{12}^\sharp e^{\bar a}$ generate non-abelian isometries with algebra
\eq{\label{iso_alg}
	\bigl[t_{12}^\sharp e^{\bar a},t_{12}^\sharp e^{\bar b}\bigr]
	&=\bigl(D^{\bar a}(t_{12})^{\bar bp}-D^{\bar b}(t_{12})^{\bar ap}
		+(t_{12})^{\bar am}\,(t_{12})^{\bar bn}\,f^p{}_{mn}\bigr)e_p\\
	&=F_{\bar m}{}^{\bar a\bar b}\,t_{12}^\sharp e^{\bar m} +R^{m\bar a\bar b}\,e_m
}
with the differential $D^{\bar a}=(t_{12})^{\bar am}e_m$.
Hence the duality isometries do not span a closed algebra in general. The defect is given by $R\in\Gamma(\bigotimes^3TM)$, which can locally be written as
\eq{\label{defect}
	R^{abc} &= (t_{12})^{bm}\partial_m(t_{12})^{ca}-(t_{12})^{cm}\partial_m(t_{12})^{ba} \\
		&\quad-\tfrac{1}{2}\bigl[(t_{12})^{ma}\partial_m(t_{12})^{bc}-(t_{12})^{ma}\partial_m(t_{12})^{cb}\bigr] \\
	&\quad+(t_{12})^{bm}\,(t_{12})^{cn}\,f^{a}{}_{mn}-(t_{12})^{ma}\,(t_{12})^{bn}\,f^{c}{}_{mn}
	-(t_{12})^{cm}\,(t_{12})^{na}\,f^{b}{}_{mn}\,.
}
With respect to this defect the structure constants $F_{\bar a}{}^{\bar b\bar c}$ can be determined in terms of the structure constants $[e_a,e_b]=f^m{}_{ab}e_m$:

\eq{\label{st_cons}
	F_{a}{}^{bc}=\tfrac{1}{2}\bigl[\partial_a(t_{12})^{bc}-\partial_a(t_{12})^{cb}\bigr]
		+(t_{12})^{bm}\,f^{c}{}_{am}-(t_{12})^{cm}\,f^{b}{}_{am}\,.
}
Thus the isometry algebra \eqref{iso_alg} closes if the defect \eqref{defect} vanishes\footnote{This condition is a priori only sufficient: Although other decompositions of $F$ and $R$ in \eqref{iso_alg} could not have been found there might be other possibilities. However, for an antisymmetric $t_{12}$ this is a natural construction.}, which is assumed in the following. This condition can conveniently be studied in terms of Lie algebroids. $t_{12}^\sharp$ maps $T^*M$ to $TM$ and can therefore be applied to general one-forms $\xi$, $\eta$: $t_{12}^\sharp\xi = \xi_{\bar a}(t_{12})^{\bar am}e_m$. Then the Lie bracket gives
\eq{\label{branch}
	\bigl[t_{12}^\sharp\xi,t_{12}^\sharp\eta\bigr] = 
	\bigl(\xi_{\bar m}\,D^{\bar m}\eta_{\bar a}-\eta_{\bar m}\,D^{\bar m}\xi_{\bar a}
		+ \xi_{\bar m}\,\eta_{\bar n}\,F_{\bar a}{}^{\bar m\bar n}\bigr)t_{12}^\sharp e^{\bar a}\,.
}
From this a Lie algebroid $(T^*M,\lb\cdot,\cdot\rb,t_{12}^\sharp)$ can be deduced. The bracket $\lb\cdot,\cdot\rb:\Gamma(T^*M)\times\Gamma(T^*M)\to\Gamma(T^*M)$ and the \emph{anchor} $t_{12}^\sharp$ have to satisfy
\begin{itemize}
\item  the Jacobi identity and the Leibniz rule $\lb\xi,f\eta\rb = f\lb\xi,\eta\rb +\bigl(t_{12}^\sharp\xi(f)\bigr)\eta$ for all $f\in\mathcal{C}^\infty(M)$,
\item the anchor property $t_{12}^\sharp\bigl(\lb\xi,\eta\rb\bigr)=\bigl[t_{12}^\sharp\xi,t_{12}^\sharp\eta\bigr]$.
\end{itemize}
As can readily be seen from \eqref{branch} and the properties of the Lie bracket, the bracket $\lb\cdot,\cdot\rb$ is given by
\eq{\label{bracket}
	\lb\xi,\eta\rb = \bigl(\xi_{\bar m}\,D^{\bar m}\eta_{\bar a}-\eta_{\bar m}\,D^{\bar m}\xi_{\bar a}
		+ \xi_{\bar m}\,\eta_{\bar n}\,F_{\bar a}{}^{\bar m\bar n}\bigr)e^{\bar a}
	\quad\quad\forall \xi,\eta\in\Gamma(T^*M)\,,
}
which fulfils the anchor property and Leibniz rule by construction.
This construction of a Lie algebroid is analogous to the one introduced in \cite{Blumenhagen:2013aia}. From the anchor property it follows that if the Lie algebroid bracket satisfies the Jacobi identity, the isometry algebra \eqref{iso_alg} satisfies it as well. It is more instructive to study the Jacobi identity for $\lb\cdot,\cdot\rb$. To this end, two cases are distinguished.
\begin{itemize}
\item\textbf{$t_{12}$ antisymmetric:} The bracket \eqref{bracket} can be written as
\eq{
	\lb\xi,\eta\rb= L_{t_{12}^\sharp\xi}\eta-\iota_{t_{12}^\sharp\eta}\mathbf{d}\xi=[\xi,\eta]_K \,,
}
i.e. it coincides with the \emph{Koszul bracket}.
It satisfies the Jacobi identity and anchor property (the anchor being $t_{12}^\sharp$) if and only if $t_{12}$ is a Poisson bi-vector; this is equivalent to the vanishing of $R$ \eqref{defect}. Hence \emph{for $t_{12}$ antisymmetric the isometry algebra \eqref{iso_alg} is a Lie algebra if and only if $t_{12}$ is a Poisson bi-vector}.
\item\textbf{$t_{12}$ symmetric:} The structure constant becomes very simple such that the Lie algebroid bracket \eqref{bracket} reduces to
\eq{
	\lb\xi,\eta\rb = \iota_{t_{12}^\sharp\xi}\mathbf{d}\eta-\iota_{t_{12}^\sharp\eta}\mathbf{d}\xi\,.
}
The Jacobi identity can be checked by using vanishing of \eqref{defect} and the Jacobi identity for the Lie bracket.
\end{itemize}
The case of an antisymmetric $t_{12}$ is of particular importance as it covers the case of $\beta$-transformations discussed in section~\ref{sec:proto}.

Now the second condition in \eqref{integ} will be discussed. Assuming $R=0$, consistency of the integrability conditions \eqref{integ} with the algebra \eqref{iso_alg} requires the two ways of evaluating $L_{[t_{12}^\sharp e^{\bar a},t_{12}^\sharp e^{\bar b}]}B$ to coincide, namely to assure
\eq{
	L_{F_{\bar m}{}^{\bar a\bar b}t_{12}^\sharp e^{\bar m}}B
	=[L_{t_{12}^\sharp e^{\bar a}},L_{t_{12}^\sharp e^{\bar b}}]B\,.
}
Using $L_{fv}B=fL_vB+df\wedge\iota_vB$ for any vector field $v$ and any function $f$, this leads to
\eq{\label{pre_rep}
	\mathbf{d}F_{\bar m}{}^{\bar a\bar b}\wedge\iota_{t_{12}^\sharp e^{\bar m}}B
		-F_{\bar m}{}^{\bar a\bar b}\,\mathbf{d}\bigl(t_{22}^\sharp e^{\bar m}\bigr)
	=-\mathbf{d}\left[L_{t_{12}^\sharp e^{\bar a}}\bigl(t_{22}^\sharp e^{\bar b}\bigr)
		-L_{t_{12}^\sharp e^{\bar b}}\bigl(t_{22}^\sharp e^{\bar a}\bigr)\right]\,.
}
This in turn is only consistent if the left-hand-side is closed, which -- using \eqref{integ} -- is equivalent to
\eq{\label{HFrel}
	\mathbf{d}F_{\bar m}{}^{\bar a\bar b}\wedge\iota_{t_{12}^\sharp e^{\bar m}}H=0\,.
}
The two immediate solutions are as follows:
\begin{itemize}
\item $F_{\bar m}{}^{\bar a\bar b}$ constant. This depends on the choice of frame $\{e_a\}_{a=1}^d$ for $TM$. Choosing a holonomic frame such as the coordinate frame, $F=0$ for $t_{12}$ symmetric. For $t_{12}$ antisymmetric, $\partial_d\partial_a(t_{12})^{bc}$ has to vanish in a holonomic frame; the components are restricted to be at most linear in the coordinates.
\item $\iota_{t_{12}^\sharp e^{\bar m}}H=0$. This is equivalent to $\iota_{t_{12}^\sharp e^{\bar m}}B+t_{22}^\sharp e^{\bar m}$ being closed. Since this requirement is not met in the simplest examples of duality (see \cite{Shelton:2005cf} or section~\ref{sec:ex}), this option will be discarded.
\end{itemize}
Although other solutions to \eqref{HFrel} are possible as well, in particular combinations of the two presented above, only the first one is applied in the following. 
For a constant $F$, the consistency condition \eqref{pre_rep} reduces up to exact terms to
\eq{\label{rep}
	L_{t_{12}^\sharp e^{\bar a}}\bigl(t_{22}^\sharp e^{\bar b}\bigr)
		-L_{t_{12}^\sharp e^{\bar b}}\bigl(t_{22}^\sharp e^{\bar a}\bigr) =
	F_{\bar m}{}^{\bar a\bar b}\,t_{22}^\sharp e^{\bar m}\,.
}
The results of this section bridge to the well-known approaches to T-duality via gauging of (multiple) dualities \cite{Hull:1989jk}. This will be discussed in the following.

\subsubsection*{The Courant algebroid for duality isometries and consistent gauging}
Above the consistency conditions on $t_{12}^\sharp e^{\bar a}$ and $t_{22}^\sharp e^{\bar a}$ have been formulated. For the former this was accomplished by the introduction of the Lie algebroid $(T^*M,\lb\cdot,\cdot\rb,t_{12}^\sharp)$. For the latter the condition \eqref{rep} has to be satisfied. Both conditions can be combined into a Courant algebroid \cite{liu1997}. The purpose for this is to bridge to the well-known approaches to T-duality via gauging of (multiple) dualities \cite{Hull:1989jk}.

It is convenient to introduce $\kappa^{\bar a}\in\Gamma(T^*M)$ given by $\kappa^{\bar a}=t_{22}^\sharp e^{\bar a} + \iota_{t_{12}^\sharp e^{\bar a}}B$. Then the dual coordinates \eqref{dual_coord} read
\eq{
	d\widetilde{X}^{\bar a} = \bigl[(t_{12})^{\bar am}G_{ma}\bigr]\star\!dX^a +\kappa_a^{\bar a}\,dX^a
}
and with $H=dB$ the integrability condition \eqref{integ} becomes
\eq{
	L_{t_{12}^\sharp e^{\bar a}}G=0 \quad\quad\mathrm{and}\quad\quad
	\iota_{t_{12}^\sharp e^{\bar a}}H=-\mathbf{d}\kappa^{\bar a}\,.
}
Evaluating the second condition for the commutator gives
\eq{
	\iota_{[t_{12}^\sharp e^{\bar a},t_{12}^\sharp e^{\bar b}]}H
	=-\mathbf{d}\bigl(L_{t_{12}^\sharp e^{\bar a}}\kappa^{\bar b}\bigr)
}
As one can see, the one-form $L_{t_{12}^\sharp e^{\bar a}}\kappa^{\bar b}$ corresponds to the vector $[t_{12}^\sharp e^{\bar a},t_{12}^\sharp e^{\bar b}]$. This suggests to combine $t_{12}^\sharp e^{\bar a}$ and $\kappa^{\bar a}$ to a generalized vector with Dorfman bracket
\eq{\label{dorfman}
	\bigl\lb t_{12}^\sharp e^{\bar a}+\kappa^{\bar a},t_{12}^\sharp e^{\bar b}+\kappa^{\bar b}\bigr\rb_D 
	= [t_{12}^\sharp e^{\bar a},t_{12}^\sharp e^{\bar b}]+L_{t_{12}^\sharp e^{\bar a}}\kappa^{\bar b}
	-\iota_{t_{12}^\sharp e^{\bar b}}\mathbf{d}\kappa^{\bar a}
		+\iota_{t_{12}^\sharp e^{\bar a}}\iota_{t_{12}^\sharp e^{\bar b}}H\,,
}
where the last two terms add-up to zero by the integrability conditions. The bracket \eqref{dorfman} is the $H$-twisted Dorfman bracket introduced in \cite{Severa:2001qm}.
In \cite{Alekseev:2004np} and more recently in \cite{Plauschinn:2013wta}, this bracket was studied in the context of isometries. Since the last two terms of \eqref{dorfman} vanish by integrability \eqref{integ} and for $R=0$, closedness of the bracket requires
\eq{
	L_{t_{12}^\sharp e^{\bar a}}\kappa^{\bar b} = F_{\bar m}{}^{\bar a\bar b}\,\kappa^{\bar m}\,;
}
then $\bigl\lb t_{12}^\sharp e^{\bar a}+\kappa^{\bar a},t_{12}^\sharp e^{\bar b}+\kappa^{\bar b}\bigr\rb_D=F_{\bar m}{}^{\bar a\bar b}(t_{12}^\sharp e^{\bar m}+\kappa^{\bar m})$.
Using the definition of $\kappa^{\bar a}$, this can be seen to agree with the consistency condition \eqref{rep} up to exact terms. Hence the closedness of the bracket \eqref{dorfman} is equivalent to closedness of the isometry algebra \eqref{iso_alg} and the consistency condition \eqref{rep}. Therefore consistency of the isometry algebra with the integrability conditions is summarized by the Courant algebroid $(TM\oplus T^*M,\lb\cdot,\cdot\rb_D,\mathrm{pr}_{TM})$.

The conventional approach to dualities is based on gauging the isometries of the sigma model \cite{Buscher:1987sk,Rocek:1991ps}. For multiple (non-abelian) isometries this procedure suffers from the introduction of anomalies \cite{Hull:1989jk}. Their absence is guaranteed if the generalized vectors $t_{12}^\sharp e^{\bar a}+\kappa^{\bar a}$ satisfy
\eq{\label{caf}
	\bigl\langle t_{12}^\sharp e^{\bar a}+\kappa^{\bar a},
		t_{12}^\sharp e^{\bar b}+\kappa^{\bar b}\bigr\rangle =
	\iota_{t_{12}^\sharp e^{\bar b}}\kappa^{\bar a}+\iota_{t_{12}^\sharp e^{\bar a}}\kappa^{\bar b}=0
}
with $\langle\cdot,\cdot\rangle$ the canonical inner product on the generalized tangent bundle.
In terms of the Dorfman bracket \eqref{dorfman}, this condition forces the subbundle spanned by $t_{12}^\sharp e^{\bar a}+\kappa^{\bar a}$ to be a Dirac structure\footnote{A Dirac structure is a maximally ($d$-dimensional) isotropic (zero inner product) and involutive (closed Dorfman bracket) subbundle of $TM\oplus T^*M$. Maximality is achieved for $t_{12}^\sharp$ surjective.}. By the duality map \eqref{dualitymap}, \eqref{dorfman} can be interpreted as the algebra of the conserved currents \eqref{dual_coord}. Then \eqref{caf} ensures anomaly freedom of the current algebra \cite{Alekseev:2004np}.

As the present approach avoids gauging the isometries, anomaly free currents and thereby the Dirac structure is not needed. In this sense, $O(d,d)$-duality requires less conditions than the conventional procedure in principle.

\subsection{Examples of $O(d,d)$-duality: The prototypes}
\label{sec:proto}
This section is devoted to examples for the duality just introduced. Beside the expected symmetries/dualities by diffeomorphisms, gauge transformations and T-duality, a novel duality induced by $\beta$-transformations will be discussed.  The coordinate frame $\{\frac{\partial}{\partial x^a}\}_{a=1}^d$ is considered for simplicity.

\subsubsection*{Coordinate transformations}
Given an invertible $d\times d$-matrix $\mathsf{A}$, the $O(d,d)$-matrix
\eq{\label{Tcoord}
	\mathcal{T}_\mathrm{diffeo} = \begin{pmatrix}
	\mathsf{A} & 0 \\ 0 & \mathsf{A}^{-t} \end{pmatrix}
}
can be considered. Applied to the generalized metric it gives
\eq{
	\mathcal{T}_\mathrm{diffeo}^t\,\mathcal{H}(G,B)\,\mathcal{T}_\mathrm{diffeo} 
	=\mathcal{H}(\mathsf{A}^tG\mathsf{A},\mathsf{A}^tB\mathsf{A})\,.
}
Therefore $\mathcal{T}_\mathrm{diffeo}$ gives rise to a change of frame of the tangent bundle. In respect of $O(d,d)$-duality, the integrability conditions \eqref{integ} are satisfied trivially and the dual coordinates are given by the change of frame $d\widetilde{X}^{\bar a} = \mathsf{A}^{\bar a}{}_{a}\, dX^a$.
Since the background transforms with the inverse, the dual action coincides with the initial one; $S(\widetilde{X};g,b)=S(X;G,B)$.
\subsubsection*{B-transformations}
Given an antisymmetric $d\times d$-matrix $\mathsf{B}$ corresponding to a two-form, a \emph{B-transformation} is given by the matrix
\eq{\label{TB}
	\mathcal{T}_\mathsf{B} = \begin{pmatrix}
	\mathds{1} & 0 \\ -\mathsf{B} & \mathds{1} \end{pmatrix}\,.
}
Conjugating the generalized metric with it results in
\eq{
	\mathcal{T}_\mathsf{B}^t\,\mathcal{H}(G,B)\,\mathcal{T}_\mathsf{B} 
	=\mathcal{H}(G,B+\mathsf{B})\,.
}
It corresponds to a gauge transformation for an exact $\mathsf{B}$, i.e. a symmetry of \eqref{sigma}. The $O(d,d)$-duality is again trivial with dual coordinate one-form $d\widetilde{X}^a = dX^a$.
Therefore the dual action becomes $S(\widetilde{X};g,b)=S(X;G,B+\mathsf{B})$.

\subsubsection*{T-duality}
Defining the $d\times d$-matrix $1_k$ by having $1$ as $(k,k)$-entry and the rest zero, the matrix
\eq{\label{TT}
	\mathcal{T}_{k} =\begin{pmatrix}
	\mathds{1}-1_k & 1_k \\ 1_k & \mathds{1}-1_k \end{pmatrix}
}
can be considered \cite{Grana:2008yw}.
From the field redefinition \eqref{redef} the components of the new metric and two-form can be determined. A tedious calculation leads to
\eq{
	g_{kk}=\frac{1}{G_{kk}}\,, \quad\quad g_{ka}=-\frac{B_{ka}}{G_{kk}}\,,
	\quad&\quad g_{ab}=G_{ab}-\frac{G_{ak}G_{kb}+B_{ak}B_{kb}}{G_{kk}} \\
	b_{ka}=-\frac{G_{ka}}{G_{kk}}\,,
	\quad&\quad b_{ab}=B_{ab}-\frac{G_{ak}B_{kb}+B_{ak}G_{kb}}{G_{kk}}
}
for $a,b\neq k$. These are the Buscher rules \eqref{buscher} in the $k^\mathrm{th}$ direction. For the integrability condition \eqref{integ} to be satisfied, the vector field $e_k$ has to be Killing with $L_{e_k}B=0$. Moreover, vanishing of \eqref{defect} and the Jacobi identity for the Killing algebra \eqref{iso_alg} are trivial for a single T-duality. Then the dual coordinate one-forms are
\eq{\label{Tcoords}
	d\widetilde{X}^k = G_{ka}\star\! dX^a +B_{ka}dX^a \quad\quad\&\quad\quad
	d\widetilde{X}^a = dX^a \quad\mathrm{for}\,a\neq k \,.
}
Hence $O(d,d)$-duality yields T-duality as a special case (cf. section~\ref{sec:Trev}). In particular, \eqref{Tcoords} coincides with \eqref{Tdual_coord}.

\subsubsection*{$\beta$-transformations}
For a antisymmetric bivector field $\beta\in\Gamma(\Lambda^2 TM)$ corresponding to an antisymmetric $d\times d$-matrix,
\eq{\label{beta-trafo}
	\mathcal{T}_\beta = \begin{pmatrix}
	\mathds{1} & -\beta \\ 0 & \mathds{1} \end{pmatrix}
}
is defined.
The transformed background \eqref{redef} induced by this \emph{$\beta$-transformation} is  given in terms of $\gamma_\beta=\mathds{1}-(G-B)\beta$ \eqref{gamma} as
\eq{\label{beta_back}
	g&= \gamma_\beta^{-1}\,G\,\gamma_\beta^{-t} \\
	b&= \gamma_\beta^{-1}\left[B-(G-B)\beta(G-B)^t\right]\gamma_\beta^{-t}\,.
}
Moreover, $O(d,d)$-duality is non-trivial: \eqref{integ} requires $\beta^\sharp e^a$ to be a Killing vector with $L_{\beta^\sharp e^a}B=0$ and consistency of the Killing algebra demands $\beta$ to be a Poisson bi-vector at most linear in the coordinates. The dual coordinate one-forms \eqref{dual_coord} are
\eq{
	d\widetilde{X}^{\bar a} = \beta^{\bar am}G_{ma}\star\! dX^a 
		+(\beta^{\bar am}B_{ma}+\delta_a^{\bar a})dX^a\,.
}
Hence, $O(d,d)$-duality establishes the classical equivalence between the sigma models $S(X;G,B)$ \eqref{sigma} and $S(\widetilde{X};g,b)$ with the coordinates and the backgrounds related by a $\beta$-transformation, provided $\beta$ is a Poisson structure and satisfies \eqref{integ}. This duality will be called \emph{Poisson duality}.
\newline

\noindent
The four particular $O(d,d)$-transformations considered here span $O(d,d)$ \cite{Blumenhagen:2013aia}. Thus, by composition non-abelian dualities are in principle covered as well. The question of conformality of $O(d,d)$-dual backgrounds will be addressed in the next section.

\subsection{Conformality of $O(d,d)$-dual backgrounds}
\label{sec:conform}
As discussed so far, $O(d,d)$-duality is a classical equivalence of constrained sigma models. For being a duality of string theory, it has to preserve conformality of the backgrounds. Change of frames \eqref{Tcoord} and exact B-transformations \eqref{TB} are symmetries and therefore retain conformality. For T-duality \eqref{TT}, the Buscher rules \eqref{buscher} have to be supplemented with a shift of the dilaton $\phi$ by $-\ln G_{kk}^2$ \cite{Buscher:1987sk,Buscher:1987qj}. By using the techniques of \cite{Blumenhagen:2013aia}, mere $\beta$-transformations, i.e. without taking a dilaton into account, can be shown to destroy conformality of an initially conformal background. In the following, \emph{$\mathcal{A}$-exact} $\beta$-transformations are argued to be a duality on the quantum-level upon an appropriate shift of the dilaton. For simplicity, $(G,B)$ is assumed to be conformal with $\phi=0$.

$\beta$-transformations can be related to B-transformations by T-duality:
\eq{\label{bchain}
	\begin{pmatrix} 0 & \mathds{1}^* \\ \mathds{1}_* & 0 \end{pmatrix}
	\begin{pmatrix}\mathds{1} & 0 \\ -\mathsf{B} & \mathds{1} \end{pmatrix}
	\begin{pmatrix} 0 & \mathds{1}^* \\ \mathds{1}_* & 0 \end{pmatrix} = 
	\begin{pmatrix}\mathds{1} & -\mathds{1}^*\mathsf{B}\mathds{1}^* \\ 0 & \mathds{1} \end{pmatrix}\,.
}
For simplicity, T-duality in every direction is considered.
The unit matrices $\mathds{1}^*$ and $\mathds{1}_*$ have to be understood in a formal manner; they act as unit on the component matrices but interchange $TM$ and $T^*M$ (cf. \eqref{odd_comp}). In particular, $\mathds{1}^*\mathsf{B}\mathds{1}^*$ is a bivector field $\sum_{a,b}\mathsf{B}_{ab}e_a\wedge e_b$. For a complete T-duality the relation between the backgrounds \eqref{redef} can be summarized as $(g+b)=\mathds{1}_*(G+B)^{-1}\mathds{1}_*$. With the Killing vectors being $\{e_a\}_{a=1}^d$, every direction has to be isometric. This is only necessary if $\beta$ has full rank. For a $\beta$ of lower rank, T-duality in the linearly independent directions is sufficient and accordingly fewer isometries are required.

The chain \eqref{bchain} of $O(d,d)$-transformations will be performed successively.
For the full T-duality to be a true duality, the dilaton has to be shifted by $-\frac{1}{2}\ln\det(G+B)$ \cite{Giveon:1994mw}. The next step in the chain \eqref{bchain} is the B-transformation. For this to be a duality, $\mathsf{B}$ has to be exact -- $\mathsf{B}=\mathbf{d}\omega$ with $\omega$ a one-form. This gives the background $\mathds{1}_*(G+B)^{-1}\mathds{1}_*+\mathbf{d}\omega$. The final background arising from the last T-duality can be written as
\eq{
	g+b= (G+B)\bigl[\mathds{1}-(G-B)\mathds{1}^*\mathbf{d}\omega\mathds{1}^*\bigr]^{-t}
	=\delta_\beta^t\,\gamma_\beta^{-t}\,.
}
By comparing with the field redefinition \eqref{redef} and \eqref{gamma}, this reproduces the correct background arising from a $\beta$-transformation \eqref{beta-trafo} with $\beta = \mathds{1}^*\mathbf{d}\omega\mathds{1}^*$. Moreover, the last T-duality induces an additional dilaton shift by $-\frac{1}{2}\ln\det[\mathds{1}_*(G+B)^{-1}\mathds{1}_*+\mathbf{d}\omega]$. Hence, the procedure just presented shows that this particular $\beta$-transformation gives dual quantum theories if the dilaton
\eq{
	\phi = -\tfrac{1}{2}\ln\det(G+B)-\tfrac{1}{2}\ln\det[\mathds{1}_*(G+B)^{-1}\mathds{1}_*+\mathbf{d}\omega]
	= -\tfrac{1}{2}\ln\det (\gamma_{\mathds{1}^*\mathbf{d}\omega\mathds{1}^*})^t
}
is introduced.

The shift of the dilaton for the full T-duality is also given by the logarithm of the determinant of $\gamma_T^t=(\mathds{1}^*)^t(G+B)$. This leads to conjecture that for $O(d,d)$-duality to be a duality on the quantum level, the dilaton has to be shifted as
\eq{\label{Tdil}
	\phi \to \phi -\tfrac{1}{2}\ln\det\gamma^t \,.
}
Changes of coordinates are an exception to \eqref{Tdil} as they do not require a shift.
The redefinition \eqref{Tdil} leaves the measure $\sqrt{|\det G|}e^{-2\phi}$, which is related to the string coupling constant, invariant. This follows from $\sqrt{|\det g|}=\sqrt{|\det G|}|\det\gamma^{-1}|$ by \eqref{redef}. A more rigorous way to derive the dilaton shift is to study the change of the path integral measure $[\mathcal{D}X]\to[\mathcal{D}\widetilde{X}]$ by \eqref{dual_coord}. In particular, up to the world-sheet operations the duality map \eqref{dualitymap} comprises $\gamma^t$, which enters the Jacobian determinant. A more detailed study is beyond the scope of this work.

\subsubsection*{Exact $\beta$-transformations}
The bivectors found above can be considered exact in the Lie algebroid $\mathcal{A}=(T^*M,$ $[\cdot,\cdot]_\mathcal{A},\mathds{1}^*)$ with bracket
\eq{
	[\xi,\eta]_\mathcal{A}=\bigl(\xi_m\,\delta^{mn}\,\partial_n\eta_a-\eta_m\,\delta^{mn}\,\partial_n\xi_a
		+\xi_m\,\eta_n\,\delta^{mp}\,\delta^{nq}\,f^k{}_{pq}\,\delta_{ka}\bigr)e^a \,.
}
The components of $\mathds{1}^*$ and its inverse $\mathds{1}_*$ are written as $\delta^{ab}$ and $\delta_{ab}$ respectively. The Lie algebroid induces a nilpotent exterior derivative on $\Gamma(\Lambda^\bullet TM)$. For a vector field $\alpha$ it reads
\eq{
	d_\mathcal{A}\alpha(\xi,\eta) = (\mathds{1}^*\xi)\,\alpha(\eta)-(\mathds{1}^*\eta)\,\alpha(\xi)
		-\alpha([\xi,\eta]_\mathcal{A})\,.
}
It follows from the anchor property $\mathds{1}^*[\xi,\eta]_\mathcal{A}=[\mathds{1}^*\xi,\mathds{1}^*\eta]$ that $\mathds{1}^*\mathbf{d}\omega\mathds{1}^*=d_\mathcal{A}(\mathds{1}^*\omega)$. Therefore, an admissible bivector $\beta$ is Poisson and of the form
\eq{
	\beta = d_\mathcal{A}\alpha \quad\mathrm{with}\quad \alpha=\mathds{1}^*\omega\,,\,\,\omega\in\Gamma(T^*M)\,,
}
and consequently exact with respect to $\mathcal{A}$.

\subsection{Poisson duality for $\mathbb{T}^3$ with $H$-flux}
\label{sec:ex}
As an easy example for Poisson duality, the flat euclidean three-torus $\mathbb{T}^3$ with
\eq{
	G=(\mathbf{d}x^1)^2+(\mathbf{d}x^2)^2+(\mathbf{d}x^3)^2
	\quad\quad\&\quad\quad
	B=h\,x^3\,\mathbf{d}x^1\wedge \mathbf{d}x^2
}
is considered.
This approximate string background is the standard toy example for discussing non-geometric backgrounds \cite{Shelton:2005cf}. Crucial in this discussion is the global structure of the background: It has to be periodic in every direction. Using $L_{fv}\xi=fL_v\xi +\mathbf{d}f\wedge\iota_v\xi$ for any vector field $v$ and due to \eqref{integ}, it turns out that the only admissible $\beta$-transformations \eqref{beta-trafo} for duality are given by constant Poisson structures with vanishing $\beta^{a3}$ and $\beta^{12}$ constant. Thus the only possibility is
\eq{
	\beta = -c\,\frac{\partial}{\partial x^1}\wedge\frac{\partial}{\partial x^2}
}
with $c\in\mathbb{R}$.
This is a trivial Poisson structure with the only non-trivial Poisson bracket being $\{x^1,x^2\}=c$. Being constant, it is $\mathcal{A}$-exact as well. The duality isometry is generated by the vectors $\beta^\sharp \mathbf{d}x^1 =\frac{\partial}{\partial x^2}$ and $\beta^\sharp \mathbf{d}x^2 =-\frac{\partial}{\partial x^1}$. These are Killing vectors for the metric $G$ and satisfy $L_{\beta^\sharp e^a}B=0$. Performing the duality, the dual pulled-back coordinate one-forms \eqref{dual_coord} read
\eq{
	d\widetilde{X}^1&=(1+chX^3)dX^1-c\star\! dX^2\,, \\
	d\widetilde{X}^2&=(1+chX^3)dX^2+c\star\! dX^1\,.
}
The new background is determined by the field redefinition \eqref{redef} and reads
\eq{\label{pdT3}
	g&=\frac{1}{c^2+(1+c\,h\,x^3)^2}\left[(\mathbf{d}x^1)^2+(\mathbf{d}x^2)^2\right] +(\mathbf{d}x^3)^2\,,\\
	b&=\frac{1}{2}\,\frac{c+2\,h\,x^3 +c(h\,x^3)^2}{c^2+(1+c\,h\,x^3)^2}\,\mathbf{d}x^1\wedge \mathbf{d}x^2\,.
}
The procedure of section~\ref{sec:conform} can be applied to this case by using a T-duality along $x^1$ and $x^2$. Hence for preserving (approximate) conformality the dilaton
\eq{\label{pddil}
	\phi = -\tfrac{1}{2}\ln\left[c^2+(1+c\,h\,x^3)^2\right]
}
has to be introduced by \eqref{Tdil}.
The following observations are made.
\begin{itemize}
\item For $c=1$, this is equivalent to the well-known Q-flux background obtained by applying T-duality in the $x^1$- and $x^2$-direction of the background $(G,B)$ with a subsequent translation\footnote{This is not a symmetry.} $x^3\to x^3-\frac{1}{h}$ \cite{Shelton:2005cf}. Going once around the $x^3$-cycle ($x^3\to x^3+1$) is a periodicity only upon applying a $\beta$-transformation to $(g,b)$. As this transformation is no symmetry of $S(\widetilde{X};g,b)$ in general, this background is referred to as being non-geometric with monodromy a $\beta$-transformation.
\item In general, the monodromy upon $x^3\to x^3+1$ for \eqref{pdT3} is given by the $O(3,3)$-matrix
\eq{\label{mono}
	\mathcal{T}_\mathrm{mono} = \left(\begin{array}{ccc|ccc}
	1-ch&0&0&0&-c^2h&0 \\
	0&1-ch&0&c^2h&0&0\\
	0&0&1&0&0&0\\\hline
	0&-h&0&1+ch&0&0\\
	h&0&0&0&1+ch&0\\
	0&0&0&0&0&1
	\end{array}\right)\,,
}
which is a combination of $\beta$- and B-transformations. This means that $x^3\to x^3+1$ gives the same background as $\mathcal{T}_\mathrm{mono}^t\mathcal{H}(g,b)\mathcal{T}_\mathrm{mono}$; thus $\mathcal{T}_\mathrm{mono}$ is the transition function for \eqref{pdT3}.
\end{itemize}
As being inequivalent to the $Q$-flux background, \eqref{pdT3} with \eqref{pddil} is an example of a new approximate non-geometric background.


\section{Conclusions}
In this paper an alternative approach to dualities based on equivalent classical backgrounds has been explored. While covering the known symmetries and T-duality, duality through $
\beta$-transformations is included as well: For $\mathcal{A}$-exact Poisson bivectors and an appropriate shift of the dilaton they are shown to provide dual backgrounds. The key ingredient besides a redefinition of the background is the relation between initial and dual coordinate one-forms.

Since the method is in principle not restricted to constant $O(d,d)$-transformations, non-abelian dualities can be treated as well. The present findings allow for decomposing them into the four generating classes -- diffeomorphisms, B-transformations, T-dualities and $\beta$-transformations. It would be interesting to study non-abelian duality more detailed in this context. Related to this, the connection to Poisson-Lie T-duality \cite{Klimcik:1995ux,Klimcik:1995jn} deserves further attention. There the condition for the existence of isometries present here is relaxed by having currents which are not conserved but obey an extremal surface condition.

Although the classical duality has been discussed to a big extent, the quantum aspects of $O(d,d)$-duality are barely studied. In particular, the conjecture for the general shift of the dilaton needs to be verified more thoroughly. Moreover, the discussion lacks a clear criterion for conformality of a dual background and in particular a criterion for the necessity of exact B-transformations. The arguments presented here rely on the symmetries and T-duality. A discussion of global aspects of the procedure from the quantum field theory point of view might be helpful.

Due to the problem of invertibility of the primitive of \eqref{dual_coord} discussed in section~\ref{sec:odd_details}, it is not clear yet whether $O(d,d)$-duality goes beyond the well-known $O(d,d;\mathbb{Z})$-duality for toroidal backgrounds. However, it avoids the procedure of gauging isometries with the associated problem of possible non-trivial holonomies and provides a direct relation between the dual coordinates via \eqref{dual_coord} (cf. \eqref{const_dual}). Moreover, all the conditions known from the conventional approach of gauging isometries are recovered and interpreted in a geometric fashion in terms of Lie and Courant algebroids. Furthermore, the approach of $O(d,d)$-duality has lead to the construction of a new (approximate) non-geometric background. Thus it seems to provide a fertile (at least) alternative approach to target-space dualities.

The conclusion will be closed with a speculation about non-commutative geometry.
Having the relation between initial and dual coordinates, it is possible to study the occurrence of non-commutative coordinates for closed strings due to dualization (see \cite{Blumenhagen:2010hj,Lust:2010iy,Blumenhagen:2011ph,Condeescu:2012sp,Mylonas:2012pg,Andriot:2012vb,Bakas:2013jwa,Mylonas:2013jha}). Given a Poisson structure $\beta$ on $T^*M$, the Poisson bracket of the coordinates is given by $\{X^a,X^b\}=(X^*\beta)(dX^a,dX^b)$. In principle this allows for computing the Poisson bracket $(\widetilde{X}^*\widetilde{\beta})(d\widetilde{X}^{\bar a},d\widetilde{X}^{\bar b})$ by using \eqref{dual_coord}. However, as the Hamiltonian remains unchanged under duality, the Poisson structure is expected to be preserved. As mentioned above, for the derivation of the equations of motion \eqref{eom:X} possible boundary terms due to winding are neglected; they are of the form
\eq{
	\int_{-\infty}^{\infty} d\tau \left[W_a\,\delta X^a\right]_{\sigma=0}^{\sigma=2\pi}\,.
}
For non-vanishing canonical winding $W_a$ at $\sigma=0,2\pi$, this gives rise to additional boundary conditions which possibly change under duality. Then a proper treatment of these might give rise to non-commutative structures analogous to the open string case (see e.g. \cite{Chu:1998qz,Seiberg:1999vs,Blumenhagen:2000fp}).


\bigskip
\noindent
\emph{Acknowledgments:}
I thank Ralph Blumenhagen, Michael Fuchs and Erik Plauschinn for discussion as well as the INFN Padua for hospitality.




\bibliography{references}  
\bibliographystyle{utphys}


\end{document}